\documentclass[pdflatex,sn-mathphys-ay]{sn-jnl}

\usepackage{graphicx}%
\usepackage{subfig}
\usepackage{multirow}%
\usepackage{amsmath,amssymb,amsfonts}%
\usepackage{amsthm}%
\usepackage{mathrsfs}%
\usepackage[title]{appendix}%
\usepackage{xcolor}%
\usepackage{textcomp}%
\usepackage{manyfoot}%
\usepackage{booktabs}%
\usepackage{algorithm}%
\usepackage{algorithmicx}%
\usepackage{algpseudocode}%
\usepackage{listings}%

\theoremstyle{thmstyleone}%
\theoremstyle{thmstyletwo}%

\theoremstyle{thmstylethree}%

\begin{document}

\title{Using Non-Linear Programming Solvers to Generate Hypothalamus-Pituitary Curves for Patients With Hypothyroidism}


\author[1]{\fnm{Robert} \sur{Petersen}}\email{peter875@purdue.edu}

\author*[1]{\fnm{Vittal} \sur{Srinivasan}}\email{srini133@purdue.edu}

\author[1]{\fnm{Stanislaw} \sur{\.Zak}}\email{zak@purdue.edu}

\author[2]{\fnm{Cary} \sur{Mariash}}\email{cmariash@iu.edu}

\affil*[1]{\orgdiv{Elmore Family School of Electrical and Computer Engineering}, \orgname{Purdue University}, \orgaddress{\street{610 Purdue Mall}, \city{West Lafayette}, \postcode{47907}, \state{Indiana}, \country{United States}}}

\affil[2]{\orgdiv{Endocrinology Division}, \orgname{Indiana University School of Medicine}, \orgaddress{\street{340 West 10th Street}, \city{Indianapolis}, \postcode{46202}, \state{Indiana}, \country{United States}}}


\abstract{Common practice in treating primary hypothyroidism is to use only Thyrothropin (TSH) to adjust the dose of thyroid replacement. In this paper, it is argued that using both TSH and free Thyroxine (FT4) values in the replacement may be beneficial in the treatment of hypothyroidism. The tool to determine the optimal value of TSH and FT4 is the Hypothalamus-Pituitary (HP) curve. These curves are models of thyroid hormone concentrations that can be used to determine patient-specific treatment strategies for individuals with hypothyroidism. By generating an HP curve for an individual with hypothyroidism, a set point is determined that represents the optimal levels of thyroid hormones in the blood. A graphical method for set point determination is proposed. A physician can then prescribe a thyroid replacement strategy to achieve this set point. In this paper, two methods for generating HP curves are proposed using non-linear programming solvers and compared with the existing methods. The proposed methods are tested using datasets from the literature, as well as measurements of patients being treated for hypothyroidism.}

\keywords{Endocrine system, Hypothalamus-Pituitary-Thyroid (HPT) axis, mathematical modeling, hypothyroidism, Hypothalamus-Pituitary curve, thyroid disease}



\maketitle
\section{Introduction}
Hypothyroidism is a condition in which the thyroid gland does not produce enough thyroid hormones, which in turn can negatively affect an individual's metabolism. Hypothyroidism has a multitude of different symptoms, including general fatigue, skin dryness, slowed speech, impaired memory, weight gain, weight loss, essentially anything that can be the result of abnormal metabolism \citep{barnes1976hypothyroidism}. This makes diagnosing the condition in patients a challenging task. In addition, two cases of hypothyroidism in different individuals can have completely different symptoms, making its diagnosis more difficult. Further complicating the condition, even once a patient has been diagnosed, the process of returning them to a healthy state can be difficult. Hypothyroidism is a surprisingly common affliction; up to 10\% of the US population is currently affected by it at clinical and subclinical levels \citet{zamwar2023}.

The Hypothalamus-Pituitary-Thyroid (HPT) axis is the interaction of three glands in the endocrine system that are responsible for regulating metabolism in humans. The axis produces and distributes key hormones, thyrotropin releasing hormone (TRH, also called TRF in some sources), thyroid stimulating hormone (TSH, also referred as thyrotropin), triiodothyronine (T3) and thyroxine (T4). 

The hypothalamus, located in the brain, produces TRH that travels to the adjacent pituitary gland. TRH in the pituitary stimulates the production and release of TSH, that flows to the thyroid gland. With TSH, the thyroid is instructed to produce T3 and T4 hormones, which will be distributed throughout the body to facilitate metabolic processes, such as regulating body temperature, an individual's heart rate, and brain development, among others. T3 and T4 concentrations in the bloodstream are detected by the pituitary and the hypothalamus, which will regulate the production of their hormones, TSH and TRH, based on the concentration of T3 and T4. TRH will cause an increase in TSH, that in turn will increase T3 and T4 levels in the bloodstream. More T3 and T4 will decrease TRH production, which in turn decreases TSH. Furthermore, unbounded T4 is also known as free T4 (FT4), and is the form of T4 that is available to tissues. FT4 occurs at predictable ratios in relation to total T4, dependent on the amount of binding proteins in blood, in about 1:1000 of FT4:T4 ratio. 

In a healthy individual, this interplay of hormones strives to reach a stable balance, called a set point, where the production and utilization of these hormones will be produced and used at a proportional rate, and their concentrations will not change. Individuals in a state of normal thyroid function, with normal levels of thyroid hormones in the blood, are called euthyroid. There are cases where individuals are not capable of reaching this set point on their own. Hypothyroidism is a primary cause of thyroid dysfunction, although cancer and other diseases can require a thyroidectomy, a partial or complete removal of a person's thyroid. In all of these cases, achievement of a euthyroid state is impossible for the individual, and their thyroid hormones must be managed therapeutically.

The current, most common, hypothyroidism treatment is prescribing a patient levothyroxine (synthetic T4) at regular intervals in order to bring the patient to the healthy reference range of T4~\citep{hueston2001treatment, CG2012}. This method has its issues, because the reference range for T4 is relatively wide and is generated from a wide cross section of patients~\citep{hadlow2013relationship}. Patients who are within the healthy range of 0.5--5 mU/L for TSH and 0.6--1.4 ng/dL for FT4~\citep{IURefRanges} do not necessarily feel better because of the treatment, and even in patients who do have symptoms abated, the process of determining the correct dose can take some time and multiple visits to a physician~\citep{goede2014novel}. In this paper, we argue that using both TSH and free Thyroxine (FT4) values in the replacement may be beneficial in the treatment of hypothyroidism.

The HP curve notion was first proposed by~\citet{goede2014hypothalamus} as a relationship between the FT4 and TSH hormones. It is unique for each individual according to a variety of biological factors, including age, sex, and general health~\cite{goede2014novel}. The HP curve is a continuous curve of TSH versus FT4 within the healthy and unhealthy ranges of both hormones. In general, a person's hormone levels will not deviate significantly from that individual's HP curve. When a thyroid function test~(TFT) is conducted to measure concentrations of T4 and TSH, one would expect to see the results of hormone values somewhere on the curve. Using the HP curve has the potential to immediately find a person's euthyroid state and help an physician determine the dosage necessary to reach that point \cite{GoedeBook}. This set point is what would provide a physician the exact concentrations of hormones to try to reach through medication, and is the motivation for developing the HP curve. In \cite{MK2024}, an analytical expression for the setpoint was derived. In addition a correlation between the physiological equilibrium described by the set point and the mathematical equilibrium of the dynamical system modeling the HPT axis was presented. In our paper, we propose a graphical method to determine the set point. We use non-linear programming solvers to generate the corresponding HP curves. It should be noted however, that the HP curve does not include T3 in its model. While T3 is an important hormone in the HPT axis, it is not included in the HP curve. This abstraction is necessary to generate the curve, but should be considered in the future when developing better methods of curve generation.


In seeking to produce a method of generating HP curves that accounts for all existing data in an individual as opposed to just two data points, in this paper, we propose methods to dynamically generate the HP curve using a model of the HPT axis and patient data. Our proposed approach should allow physicians to provide better treatment to individuals with hypothyroidism.

The contributions of the paper are:
\begin{enumerate}
    \item A method is proposed to dynamically  generate HP curves for euthyroid and hypothyroid individuals that uses a non-linear programming solver to determine parameters of the HPT axis model.
    \item The HPT axis model proposed by~\cite{bonfantiendocrine} is extended and modified by estimating the constants in the Michaelis-Menten equation as parameters instead of assuming them to be fixed. This idea results in more accurate HP curves.
    \item A comparative study of the proposed and existing methods for finding the HP curves for individuals is presented. The proposed model in this paper and models from literature are verified using existing datasets and new patient data.
    \item The HP curves are generated for patients with hypothyroidism as well as from patient data in the literature. These curves can be used as a tool to aid physicians to prescribe the correct dosage of medication to bring the patients to the desired set point.
\end{enumerate}

This paper is structured as follows. In Section~\ref{sec:model}, we present a mathematical model of the HPT axis that we use to derive the HP curves for healthy individuals and patients with hypothyroidism. In Section~\ref{sec:ParamEst}, we develop an algorithm for parameter estimation of the HPT model and dynamic generation of HP curves.  In Section~\ref{sec:valid}, we validate the proposed algorithm with numerical simulations using real patient data. Finally, in Section~\ref{sec:conc}, we present conclusions and discuss open problems.

\section{HPT Axis Mathematical Model}
\label{sec:model}
In this section, we present a mathematical model of the HPT axis that we use to generate the HP curve. We require a model of the HPT axis that uses the concentrations of TSH and FT4 and that allows us to fit the model parameters to patient data. With this, we will be able to estimate all values of the hormone relationship, which, in turn, would allow us to develop an HP curve. This will allow us to use all the available data in an individual to generate the HP curve.

In this paper, we use the HPT axis model proposed by~\cite{bonfantiendocrine} which has the form,
\begin{equation}
   \left. \begin{array}{cll}
        \frac{d\left[TSH\right]}{dt} &=& k_1 - \frac{k_1\left[FT_4\right]^{n_1}}{k_a + \left[FT_4\right]^{n_1}} - k_2 \left[TSH\right]\\
        \frac{d\left[FT_4\right]}{dt} &=& k_5 - k_4\left[FT_4\right]\end{array}\right\}
\label{eq:HP_curve_model}
\end{equation}
where $k_1$ is the TSH maximum secretion rate constant of TSH, $k_2$ and $k_4$ are the degradation rates of TSH and FT4, respectively, $k_a$ and $n_1$ are Michaelis-Menten enzyme kinematics constants, $k_5$ is a constant that incorporates TSH and thyroid contributions which are assumed to be unchanging in the equation, and $\left[TSH\right]$, $\left[FT4\right]$ are the concentrations of TSH and FT4 in the bloodstream. In all cases, the constants must be positive in order to produce meaningful physiological results~\citep{kyrylov2005}. The parameter values shown in Table~\ref{tab:parameters} come from~\citep{bonfantiendocrine}.

\begin{table}[h]
\caption{Parameter values in model in~\eqref{eq:HP_curve_model} used in~\cite{bonfantiendocrine}.}
\label{tab:parameters}
\begin{tabular}{@{}lll@{}}
\toprule
Parameter & Value  & Unit            \\
\midrule
$k_1$     & 5000   & mU/(L·day)       \\
$k_4$     & 0.099  & 1/day            \\
$n_1$     & 4      & (dimensionless)  \\
\bottomrule
\end{tabular}
\end{table}

The parameters that must be estimated are $k_a$, $k_2$, and $k_5$. However,~\cite{bonfantiendocrine} argues that the term $n_1$ can be held as a constant due to its relatively consistent estimation at the value 4. Because $n_1$ is an exponent, even slight variations in it can have dramatic effects on the performance of the model. We also note that the parameter $n_1$ cannot be kept constant at $n=4$ for a general model as the exponent of the Michaelis-Menten equation may change based on each individual hormone levels. For these reasons, we will also estimate the parameter $n_1$ in~\eqref{eq:HP_curve_model}, so as to account for the variations in $n_1$ that influence the overall model. The method for this parameter estimation is described in Section~\ref{sec:ParamEst}.


With an accurate HP curve, the euthyroid state can be determined as a set point, the spot on the curve with maximum curvature. At this point, there is the smallest sensitivity to changes in the levels of either hormone. This would make sense biologically, because when hormone levels are near the desired state of an individual, small changes in either hormone concentration would not dramatically affect the concentration of the other. Conversely, at the extreme ends of the curve, small changes in the independent hormone would result in a large change in its counterpart. Recall, the set point is a point at which a euthyroid individual's hormones are in balance.



Thus, the resultant system has two equations and four unknown parameters $k_2$, $k_4$, $k_a$, $n_1$. From this, a curve can be generated using data points taken from patients that allow for approximation of the parameters. The method of accomplishing this parameter estimation is discussed in the following section.

\section{HPT Axis Model Parameter Estimation} \label{sec:ParamEst}
In this section, we present methods for estimating parameters of the HPT axis model given by~\eqref{eq:HP_curve_model}. We begin by presenting three methods that will be used to generate the HP curves. We first present the method of~\cite{goede2014novel}, followed by two novel methods. These methods will be tested against each other for accuracy, as well as against patient data sets.

\subsection{Method of~\cite{goede2014novel}} 
\begin{figure}
    \centering
    \caption{Different possible HP curves generated with~\eqref{eq:Algeb_coeff} using the same data set.}
    \includegraphics[width=.67\linewidth]{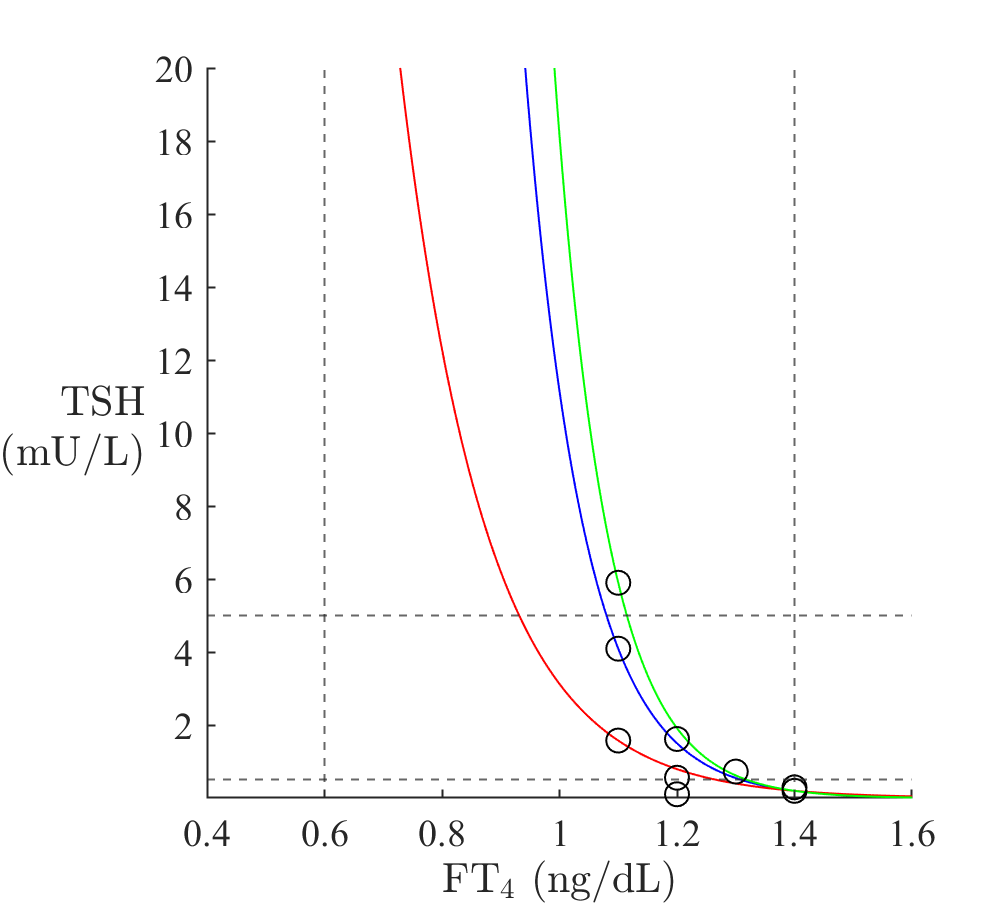}
    \label{fig:GoedeBreaker}
\end{figure}

Goede et. al described a method to construct the HP curve and set point using any pair of two patient data points. The authors use a logarithmic equation to generate parameters for an equation, which can then generate data points for a curve. The equation has the form,
\begin{equation} \label{eq:GoedeEstimator}
    [\text{TSH}] = S\exp(-\varphi[\text{FT4}]),
\end{equation}
where \textit{S} is the multiplier of the exponential equation and \textit{$\varphi$} is the slope of the exponential coefficient. The values of these two parameters are established using patient data and are given by
\begin{equation} 
\label{eq:Algeb_coeff}
\left.
\begin{split}
      \varphi &= \bigg(\frac{1}{([\text{FT4}]_1 - [\text{FT4}]_2)} \bigg)\ln\bigg({\frac{[\text{TSH}]_2}{[\text{TSH}]_1}}\bigg)\\
      S &= [\text{TSH}]_1\exp(\varphi[\text{FT4}]_1)
\end{split}
\right\}
\end{equation}
where the values of FT4 and TSH are paired data points taken from patients. Any two points can be used to generate the curve. While this method can generate HP curves, it is subject to the constraint that the two data points must be on or near the individual's actual HP curve. This is not always achievable in the clinical setting. Fig.~\ref{fig:GoedeBreaker} demonstrates this constraint; with the same data set, three different curves can be generated. This demonstrates an issue with this method. Any reading of hormone levels could contain an outlier, and if the data point used to generate the HP curve is an outlier, it could generate an inaccurate curve. In some cases, this method could also produce false healthy HP curves that indicate that a patient has a healthy set point. Although most of the patient data points could indicate unhealthy values, if the patient has two outliers in the healthy range, this could produce an HP curve with a healthy set point. Hence, this method requires choosing the best data points and eliminating any potential outliers in a given dataset.

\subsection{Non-linear Least Squares Method (NLSM)}
\cite{bonfantiendocrine} used~\eqref{eq:HP_curve_model} to generate the trajectory of an individual in the state plane FT4-TSH over several days in order to track the return to their set point. To do this,~\cite{bonfantiendocrine} used a nonlinear least squares method applied to a data set to estimate the parameters of the HPT axis given by~\eqref{eq:HP_curve_model}.

In our proposed approach, we consider the HPT axis model~\eqref{eq:HP_curve_model} in their equilibrium states. This is obtained by setting the derivatives equal to zero, resulting in two algebraic equations,
\begin{eqnarray}
    \text{[FT4]} &=& \frac{k_5}{k_4}\\
    \label{steadyFt4}
    \text{[TSH]} &=& \bigg(k_1 - \frac{k_1\text{[FT4]}^{n_1}}{k_a+\text{[FT4]}^{n_1}}\bigg)\frac{1}{k_2}
\label{eq:steadyTSH}
\end{eqnarray}
We now proceed to estimate parameters $k_a$, $k_2$, and $n_1$ iteratively using a nonlinear least squares method (NLSM). We use the objective function
\begin{equation}
\text{NLSM} = \|\text{[TSH]}_{\text{curve}} - \text{[TSH]}_{\text{data}}\|^2
\label{eq:bonfantiMinimization}
\end{equation}
where, [TSH]$_{\text{data}}$ is a vector of the real patient data, and [TSH]$_{\text{curve}}$ is a vector of the TSH concentration calculated by solving~\eqref{eq:steadyTSH} with the corresponding patient FT4 values and the estimated parameters. Iterating through this method multiple times with different parameters will decrease the values of the objective function. When the program finds what it believes to be the minimum possible value of the objective function~\eqref{eq:bonfantiMinimization}, it returns the optimal parameters that minimize the objective function NLSM. Then, using the estimated parameters and an array of the range of possible values of FT4, we use~\eqref{eq:steadyTSH} to generate an estimated HP curve. Note that this approach does not depend on the estimated value of FT4, so for that reason $k_5$ does not need to be estimated.

\subsection{ Graphical method of determining the setpoint}
\begin{figure}[h!]
    \centering
    \includegraphics[width=1\linewidth]{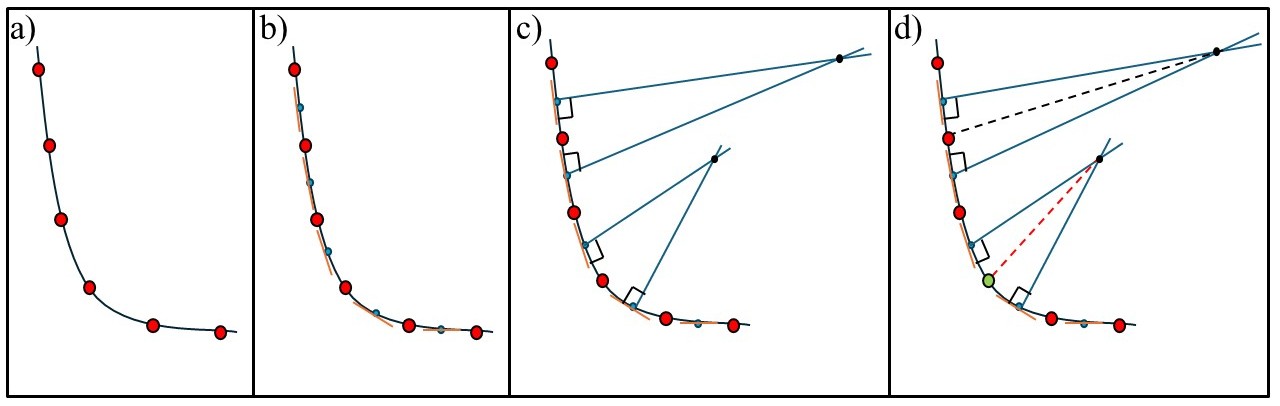}
    \caption{Illustration of the graphical method of determining setpoint. In fig.~\ref{fig:radius}(a), we select a random set of equally spaced points (red dots) on the HP curve. In fig.~\ref{fig:radius}(b), we choose two neighbors to each point and draw tangents of the neighbor points on the curve. In \ref{fig:radius}(c), the normals are constructed from the midpoint of the tangents of the neighbors for two chosen points on the HP curve for clarity. This process is repeated for all the chosen points on the HP curve. Fig.~\ref{fig:radius}(d) shows the point of maximum curvature, i.e., the setpoint, (green dot) as it has the least distance from the intersection of the corresponding normals.}
    \label{fig:radius}
\end{figure}

Having obtained the HP curve, we can find the set point, which is the point of maximum curvature on the HP curve. Maximum curvature is the point on a curve that has the smallest inside radius. The methods proposed in ~\cite{goede2014novel} and in \cite{MK2024} are analytical methods for finding the maximal curvature. We propose a graphical method for finding the set point. We select a set of points on the HP curve and then determine maximum curvature as follows:
\begin{enumerate}
    \item Select a random set of equally spaced discrete points on the HP curve.
    \item Starting from left to right, for each point, choose two neighbors to the point and draw tangents of the neighbors on the curve.
    \item Construct the normal from the midpoint of the tangents drawn previously and find the intersection of the normal lines.
    \item Find the distance between the intersection and the selected point on the HP curve.
    \item Repeat this procedure for all points on the HP curve and find the point on the HP curve with the shortest distance between the point and intersection of the corresponding normal lines. This is the point of maximum curvature of the HP curve.
\end{enumerate}
We illustrate the above algorithm of determining the set point in Fig.~\ref{fig:radius}.  

\subsection{Nonlinear Programming Solver (NLPS) Method}
We used the non-linear programming solver of MATLAB to estimate the parameters of the HPT axis model given by~\eqref{eq:HP_curve_model}. This approach uses MATLAB's \texttt{fmincon} function. We test three different objective functions. The first objective function is the Root Mean Square Error (RMSE), given by
\begin{equation} \label{eq:RMSE}
    \text{RMSE} = \sqrt{\frac{1}{n} \sum_{i=1}^{n} \left( \text{[TSH]}_{\text{curve,i}} - \text{[TSH]}_{\text{data,i}} \right)^2 }
\end{equation}
where $n$ is the number of patient data samples used. The second objective function is the Sum of Absolute Error (SAE), given by
\begin{equation} \label{eq:SAE}
{\text{SAE}} = \sum_{i=1}^{n} \left|\text{[TSH]}_{\text{curve,i}} - \text{[TSH]}_{\text{data,i}}\right|
\end{equation}
The third objective function is the Huber loss function (HLF), given by 
\begin{equation} \label{eq:Huber}
{\text{HLF}} =
    \sum_{i=1}^{n}
\begin{cases} 
\frac{1}{2} (\text{[TSH]}_{\text{curve}} - \text{[TSH]}_{\text{data}})^2 & \text{if } |\Delta| \leq \delta \\
\delta \left(\text{[TSH]}_{\text{curve}} - \text{[TSH]}_{\text{data}} - \frac{\delta}{2} \right) & \text{if } |\Delta | > \delta
\end{cases}
\end{equation}
where $\delta$ is a threshold that determines whether each data point should be subject to a more sensitive objective function if it is near the expected HP curve, or less sensitive objective function if the data point is expected to be an outlier. The difference between the TSH value of the estimated curve and the TSH value of the patient is $\Delta$.

The function \texttt{fmincon} performs constrained optimization subject to constraints. The constraints in our case are
\begin{align}
    &x>0   \\
    \alpha<&x\leq \beta
\label{constraints}
\end{align}
where $x$ is a vector of the parameters to be estimated, $\alpha$ and $\beta$ are the upper and lower bounds of the parameters, which are arbitrarily set at 0 and 1000 respectively.

\subsection{Perfomance Comparison of NLPS for Three Objective Functions}

The performance of the NLPS method with the three objective functions was compared with the performance of the method of~\cite{goede2014novel} serving as a benchmark. All three objective functions were minimized using the NLPS method, and the goodness of fit was calculated using the  $R^2$ performance indicator given by
\begin{equation} \label{eq:rsquare}
    R^2 = 1-\frac{\text{Sum squared regression (SSR)}}{\text{Total sum of squares (SST)}}
\end{equation}
where the numerator is the sum of residuals; the difference between the expected point (the HP curve of Goede et al.) and the actual point, (the HP curve generated by the NLPS method), squared. The denominator is the distance of the estimated model away from the mean of the Goede et al. model, squared. A higher $R^2$ value indicates a better performance of the method.

\begin{figure}[t]
    \centering
    \includegraphics[width=.67\linewidth]{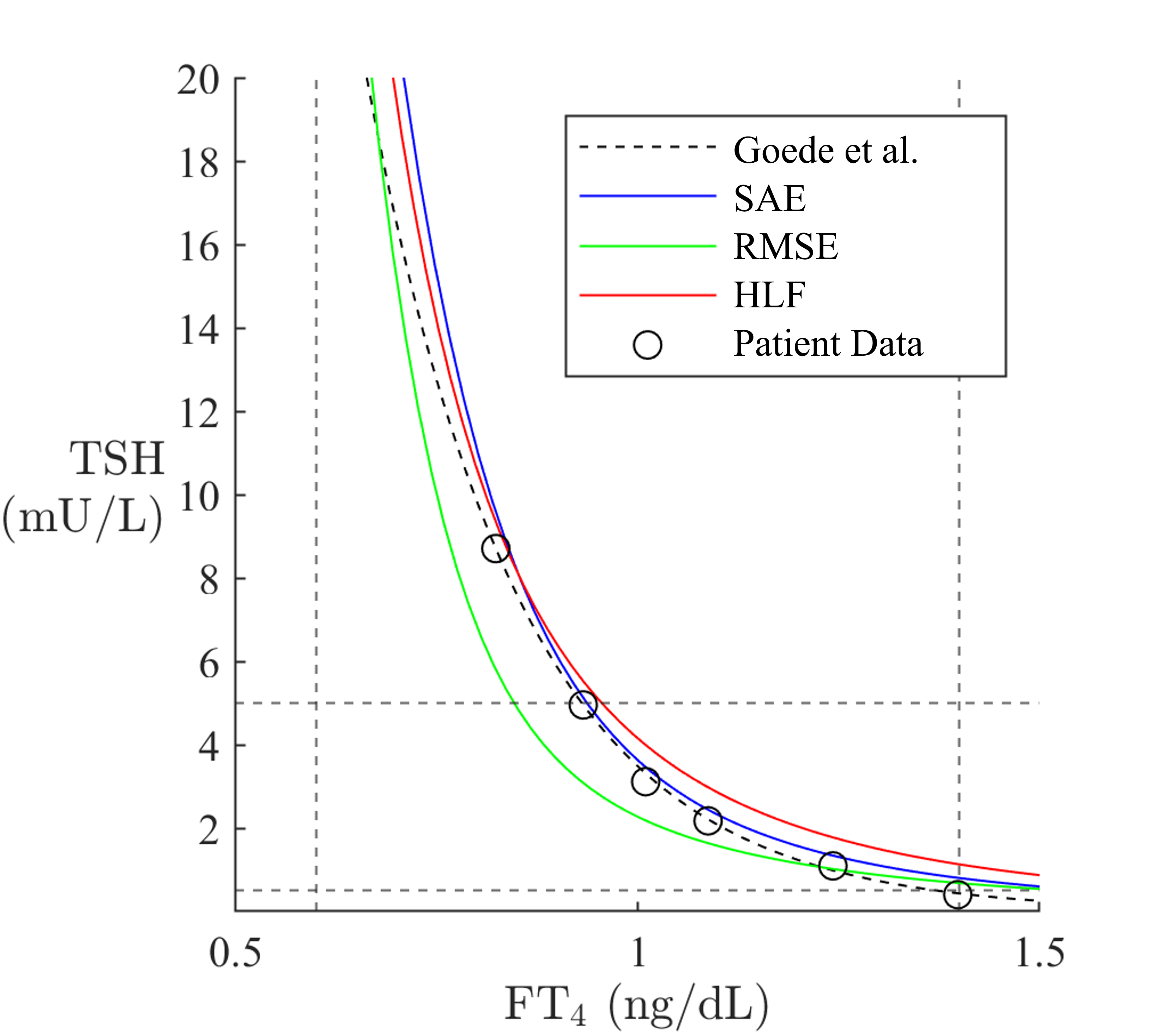}
    \caption{Objective function performance using Patient 3G from \cite{goede2014novel} for the NLPS method.}
    \label{fig:costFunctions}
\end{figure}

Fig.~\ref{fig:costFunctions} shows the comparative performance of the objective function optimization for the three methods using the~\cite{goede2014novel} data set presented in Table~\ref{tab:ft4_tsh}. All of the described models carry characteristics of the HP curve. At low levels of FT4, TSH is appropriately high, as would be expected in a euthyroid patient. Similarly, at high levels of FT4, TSH is low. In all of the curves, there is a definite point of curvature within the reference range of hormone levels, which is expected as the set point would normally be within the reference ranges. 

\begin{table}[b]
\caption{$R^2$ values for objective functions.}
\label{tab:R2values}
\begin{tabular}{@{}lccc@{}}
\toprule
Patient     & SAE    & RMSE   & HLF    \\
\midrule
Patient 1G  & 0.8847 & 0.6737 & 0.9584 \\
Patient 2G  & 0.9780 & 0.9271 & 0.9800 \\
Patient 3G  & 0.9784 & 0.7621 & 0.9494 \\
Patient 4G  & 0.9399 & 0.9076 & 0.9434 \\
\bottomrule
\end{tabular}
\end{table}

Table~\ref{tab:R2values} shows the resulting R$^2$ values. The SAE and HLF functions closely mirror the patient data and the method of~\cite{goede2014novel} The RMSE method has the poorest performance of the three methods. The HLF method routinely performs well, it has the best or near the best performance, even when the other methods do significantly worse. The HLF function's consistency in delivering a good result is why we select it to be the objective function for the NLPS in the next section.

\subsection{Advantages and Benefits of NLPS Method}
In contrast to the method of~\cite{goede2014novel}, this optimization method has the benefit of accounting for all available data points and generating a curve based on the entirety of the data.

A benefit of the NLPS method over the NLSM is that the NLSM method is by nature more sensitive to outliers and data that deviates from the individual's actual set point. Because it will generally try to minimize error across all data points, it can sometimes sacrifice the overall accuracy of the HP curve in exchange for a lower cost of the objective function. With data that perfectly matches an HP curve the NLSM performs well, but when uncertainty or noise is present, it can degrade the final proposed curve. The NLPS method is less sensitive to measurement noise. It can account for outliers and noisy data, which allows for more datapoints to be used and ultimately a more accurate HP curve.

\section{Comparative Study of NLPS and NLSM methods} \label{sec:valid}
In this section, we present the HP curves for different scenarios that are generated on real patient data sets comparing the results of the method of~\cite{goede2014novel}, the proposed improved NLSM, and the proposed HLF-NLPS methods. We solve~\eqref{eq:steadyTSH} using the optimization implemented using the \textbf{fmincon} function of MATLAB. We first compare the existing NLSM method adapted from \cite{bonfantiendocrine} where the parameter $n_1$ from~\eqref{eq:HP_curve_model} is treated as a constant, and the improved NLSM method where $n_1$ is estimated, with the~\cite{goede2014novel} method as a benchmark. Next, we generate the HP curves for eight individuals, four of whom have hypothyroidism and are on thyroid replacement therapy, as well as four individuals from~\cite{goede2014novel}. We then analyze the obtained HP curves.

\subsection{Dataset}
For data, we used data points given in~\cite{goede2014novel}, as well as data sets compiled from hypothyroid patients during periodic physical examinations over several years at Indiana University. The data taken from~\cite{goede2014novel} are included here in Table~\ref{tab:goede_data}. 
The data points for FT4 in Table~\ref{tab:goede_data} were converted to ng/dL from pmol/L to be consistent with the Indiana University data sets. The conversion rate used was

\begin{equation}
    0.07768 \; \mbox{pmol/L} =1 \mbox{ ng/dL}.
\end{equation}
Indiana University data sets were taken from anonymous patients known to have primary hypothyroidism and were undergoing treatment with oral levothyroxine replacement. The patient data sets are shown in Table~\ref{tab:goede_data} and Table~\ref{tab:ft4_tsh}.

\begin{table}[h]
\caption{Datasets with FT4 (pmol/L) and TSH (mU/L) from \cite{goede2014novel}.}
\label{tab:goede_data}
\begin{tabular}{@{}llllllll@{}}
\toprule
\multicolumn{2}{l}{\textbf{Patient 1G}} & \multicolumn{2}{l}{\textbf{Patient 2G}} & \multicolumn{2}{l}{\textbf{Patient 3G}} & \multicolumn{2}{l}{\textbf{Patient 4G}} \\
FT4 & TSH & FT4 & TSH & FT4 & TSH & FT4 & TSH \\
\midrule
1.79 & 0.05 & 3.34 & 0.01 & 0.82 & 8.71 & 0.68 & 6.30 \\
0.85 & 7.45 & 0.85 & 6.21 & 0.93 & 4.96 & 0.60 & 9.60 \\
1.17 & 1.71 & 1.17 & 3.25 & 1.24 & 1.10 & 0.85 & 1.40 \\
0.62 & 25.33 & 1.24 & 1.74 & 1.09 & 2.18 & 0.52 & 6.30 \\
1.01 & 0.05 & 3.03 & 0.00 & 1.01 & 3.12 &      &      \\
1.17 & 0.89 & 0.70 & 0.02 & 1.40 & 0.41 &      &      \\
1.55 & 0.04 & 0.78 & 0.87 &      &      &      &      \\
     &      & 1.01 & 4.97 &      &      &      &      \\
\botrule
\end{tabular}
\end{table}

\begin{table}[h]
\caption{Datasets with FT4 (ng/dL) and TSH (mU/L) taken from the University of Indiana.}
\label{tab:ft4_tsh}
\begin{tabular}{@{}llllllll@{}}
\toprule
\multicolumn{2}{l}{\textbf{Patient 1IU}} & \multicolumn{2}{l}{\textbf{Patient 2IU}} & \multicolumn{2}{l}{\textbf{Patient 3IU}} & \multicolumn{2}{l}{\textbf{Patient 4IU}} \\
FT4 & TSH & FT4 & TSH & FT4 & TSH & FT4 & TSH \\
\midrule
1.2 & 0.03 & 1.1 & 5.9 & 1.4 & 0.01 & 2.57 & 59 \\
2.27 & 0.01 & 1.2 & 1.62 & 1.0 & 1.31 & 33.47 & 0.7 \\
1.43 & 7.34 & 1.3 & 0.73 & 1.1 & 0.31 & 7.72 & 43 \\
1.1 & 1.08 & 1.2 & 0.56 & 1.4 & 0.04 & 14.16 & 2.9 \\
2.12 & 0.54 & 1.2 & 0.11 & 1.2 & 0.03 & 16.74 & 15 \\
1.5 & 11.8 & 1.4 & 0.3 & 1.0 & 0.2 & 11.59 & 46 \\
1.1 & 4.15 & 1.4 & 0.2 & 1.2 & 0.08 & 23.17 & 3.2 \\
1.2 & 0.31 & 1.1 & 4.1 & 0.9 & 7.95 & 11.59 & 2.4 \\
      &      & 1.1 & 1.58 & 0.7 & 18.35 & 5.15 & 53 \\
      &      &      &      & 0.8 & 34.10 &       &       \\
\botrule
\end{tabular}
\end{table}

Of the four patients, Patient 1IU had numerous outlier data points in the reported TSH vs FT4 values. It is unclear whether this was a result of an error in clinical measurement or an abnormality in the hormone levels of Patient 1IU. The points of the other patients were reasonable. FT4 and TSH have a negative exponential relationship, where low values of one hormone correspond to high levels of the other. Patient 2IU's data notably have several values for TSH with the same FT4 value, giving the appearance of ``stacked" data points, which can make curve fitting difficult. This can possibly be accounted for because of the lower sensitivity of the FT4 measurements, where slight variations in actual FT4 concentrations are all represented as the same level of FT4 in the measurement. Patient 4IU represents an extreme case of hypothyroidism, where the individual had FT4 levels so low that they had to be hospitalized for it.

\subsection{Comparison of the proposed NLSM method versus NLSM method from~\cite{bonfantiendocrine}}
\begin{figure}[t]
    \centering
    \includegraphics[width=.67\linewidth]{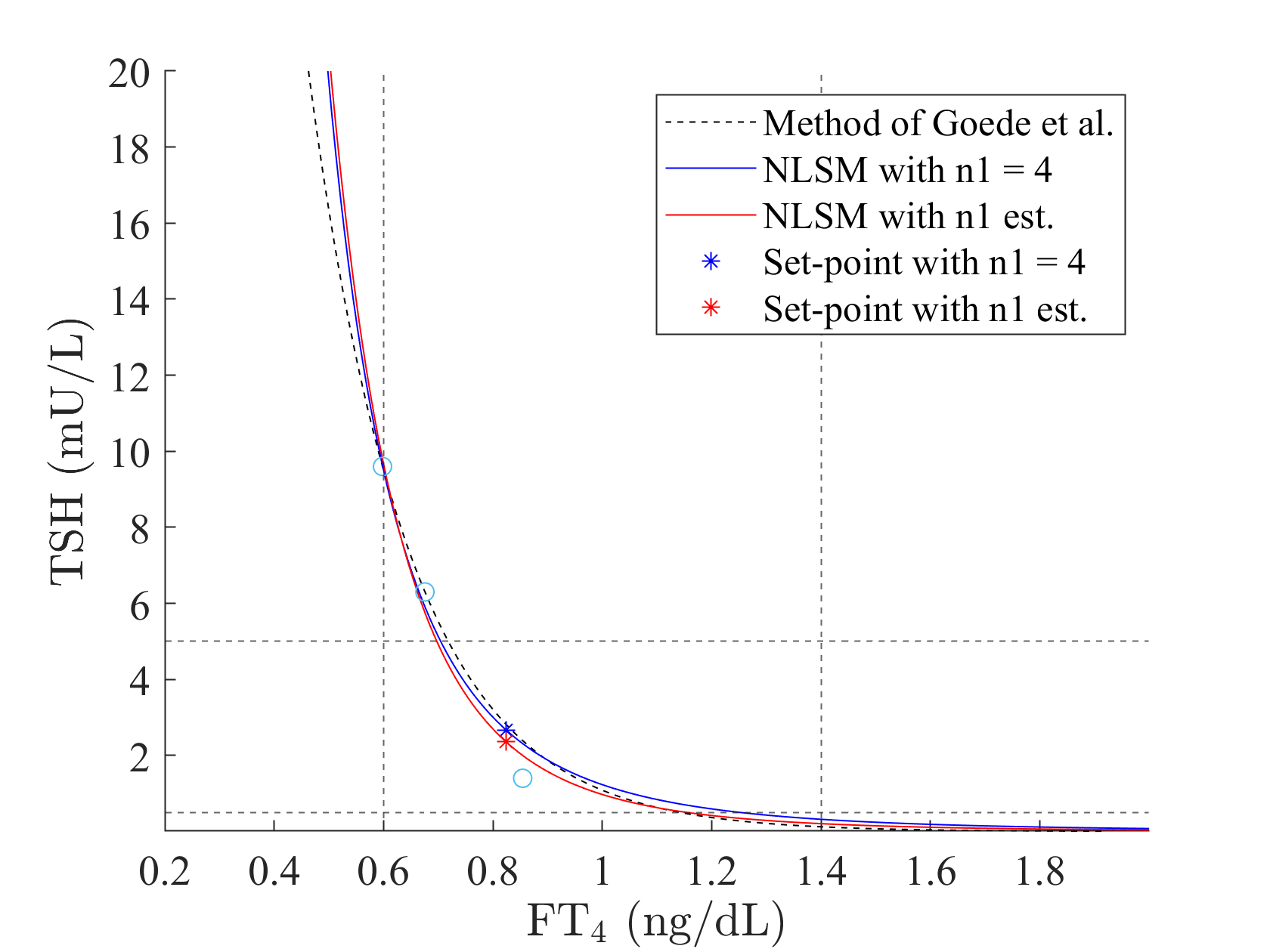}
    \caption{Comparison of proposed NLSM vs original NLSM for Patient 3G from Table~\ref{tab:goede_data}.}
    \label{fig:3N4}
\end{figure}

Fig.~\ref{fig:3N4} shows a comparison of two approaches to solving the least squares method, where the parameter $n_1$ is estimated and one where it is fixed at $n_1$ = 4. The method of Goede et al.~\cite{goede2014novel} was used to plot the HP curve and served as a benchmark for both methods. The simulation used data supplied by~\cite{goede2014novel}. We see that both methods produce curves that reflect the~\cite{goede2014novel} curve. Table~\ref{tab:NLSMNormdifferences} shows the normalized difference in set points between the estimated and fixed exponent $n_1$. The curve using an estimation of $n_1$ is generally closer to the curve of~\cite{goede2014novel} set point than the curve corresponding to the fixed $n_1$. The exception to this is in Patient 2G of the Goede et al. data set, but in this case the difference in the two set points is minimal, and both set points are closer to the benchmark than in any other data set. The method involving estimation of $n_1$ performs better than the method using a fixed $n_1$. For this reason, we will use the method that approximates $n_1$ to generate HP curves in our simulations.

\begin{table}[h]
\caption{Normalized distance between set points of NLSM using the Goede et al. data set from Table~\ref{tab:goede_data}.}
\label{tab:NLSMNormdifferences}
\begin{tabular}{@{}lcc@{}}
\toprule
Patient     & $n_1 = 4$ & $n_1 \neq 4$ \\
\midrule
Patient 1G  & 1.405     & 1.177        \\
Patient 2G  & 0.303     & 0.435        \\
Patient 3G  & 1.231     & 0.736        \\
Patient 4G  & 0.931     & 0.772        \\
\bottomrule
\end{tabular}
\end{table}


\subsection{Patient Data Verification}
\begin{figure*}[t!]
\centering
\subfloat[Patient 1G]{\includegraphics[width=0.33\textwidth, keepaspectratio]{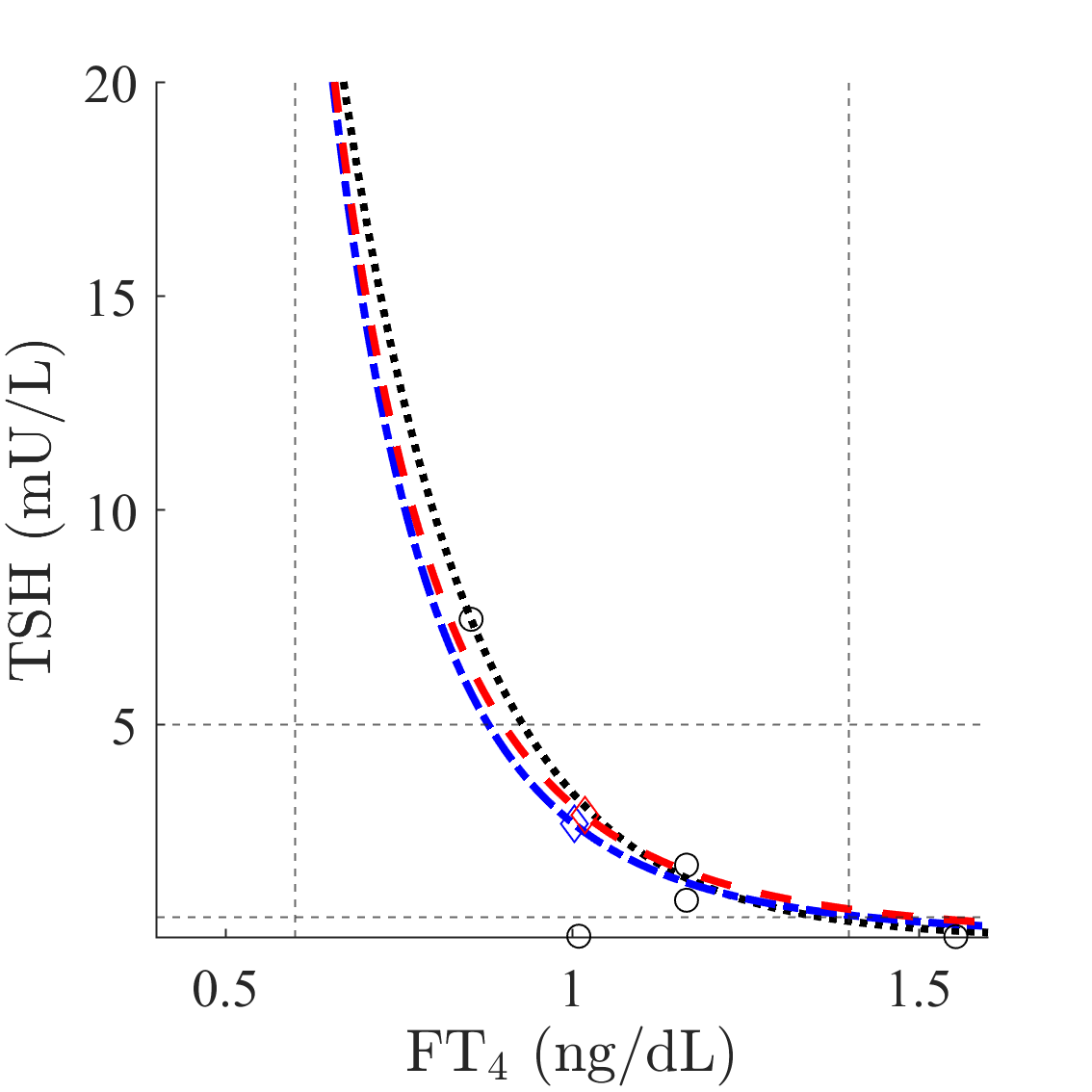}\label{fig:goede1}}
\subfloat[Patient 2G]{\includegraphics[width=0.33\textwidth, keepaspectratio]{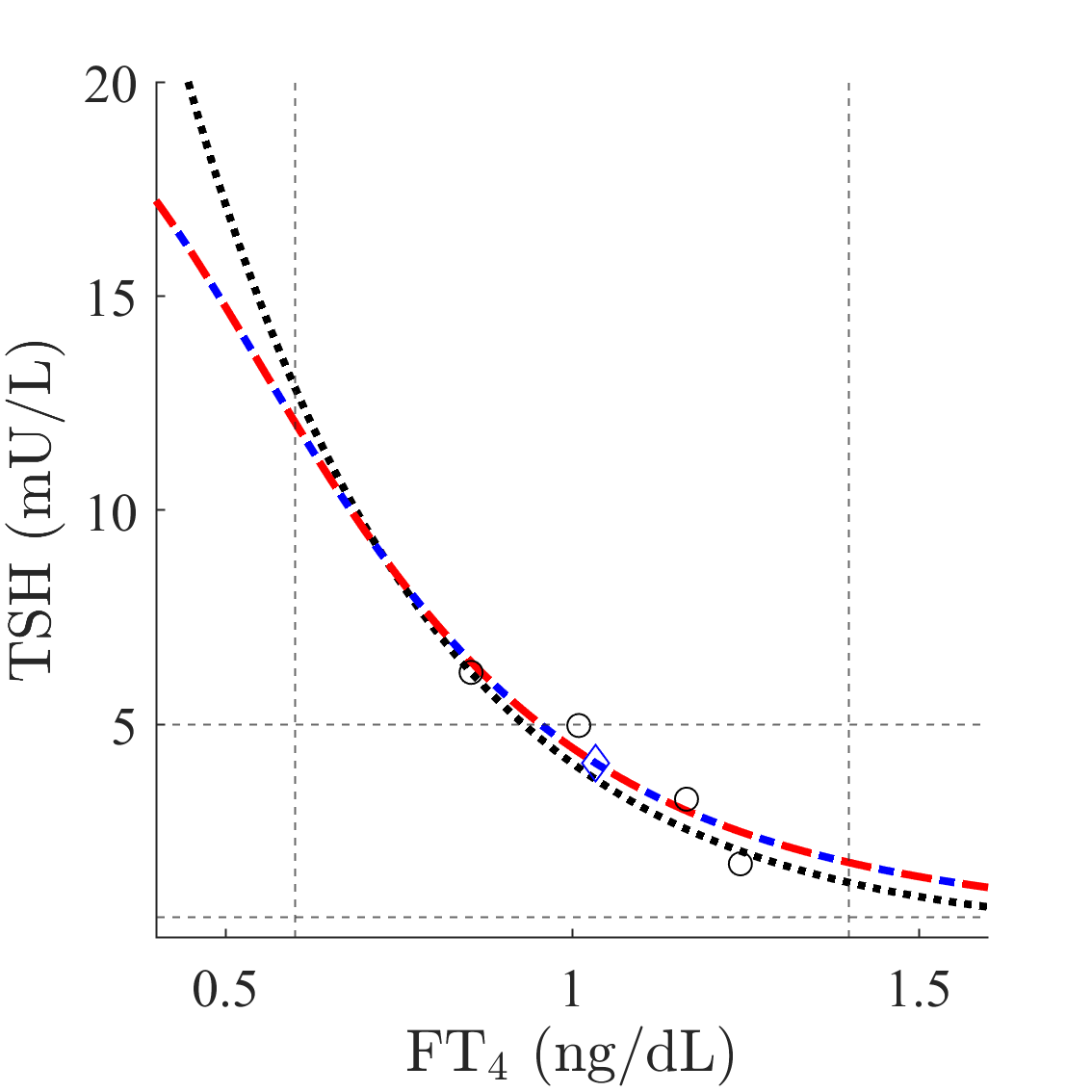}\label{fig:goede2}}
\subfloat[Patient 3G]{\includegraphics[width=0.33\textwidth, keepaspectratio]{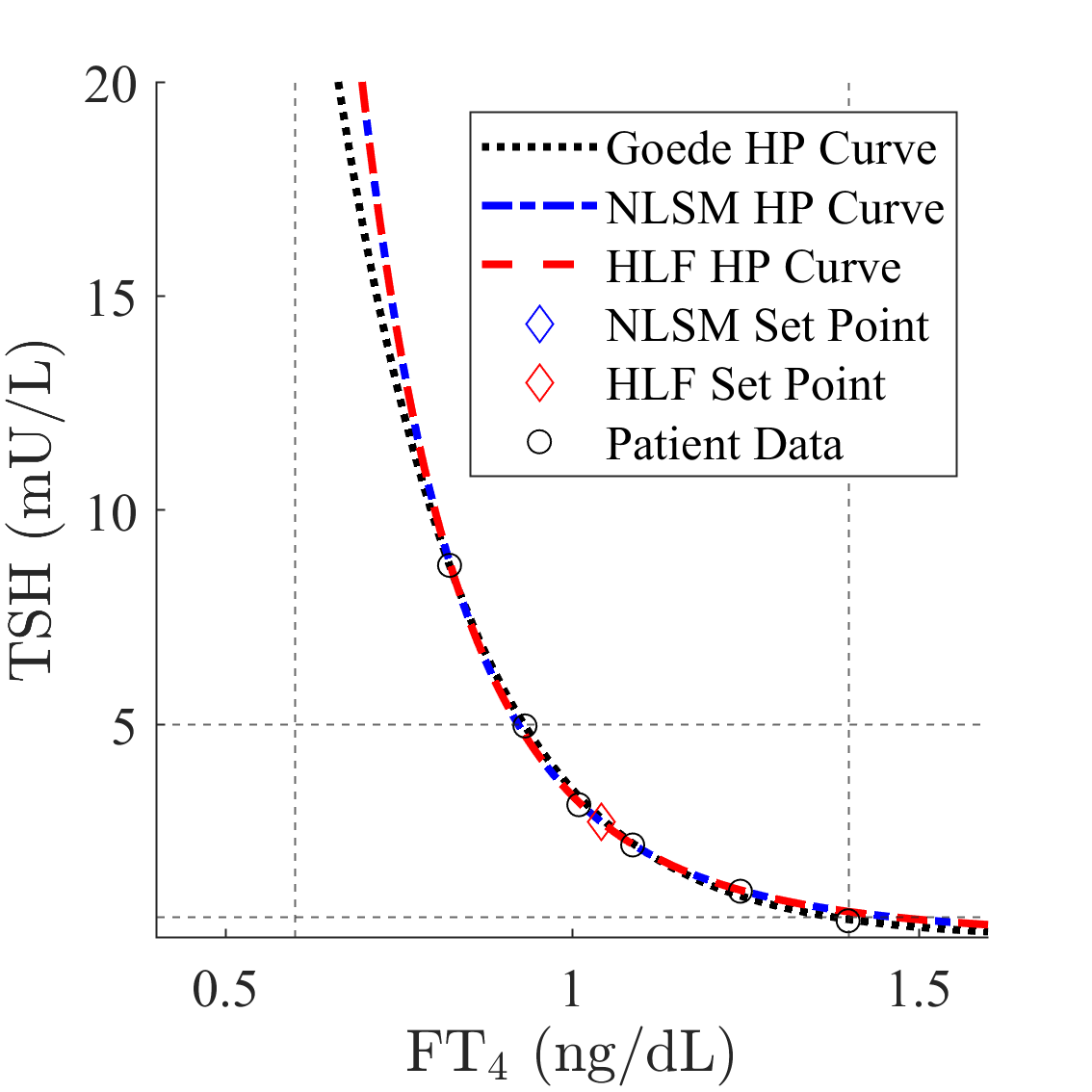}\label{fig:goede3}}

\subfloat[Patient 4G]{\includegraphics[width=0.33\textwidth, keepaspectratio]{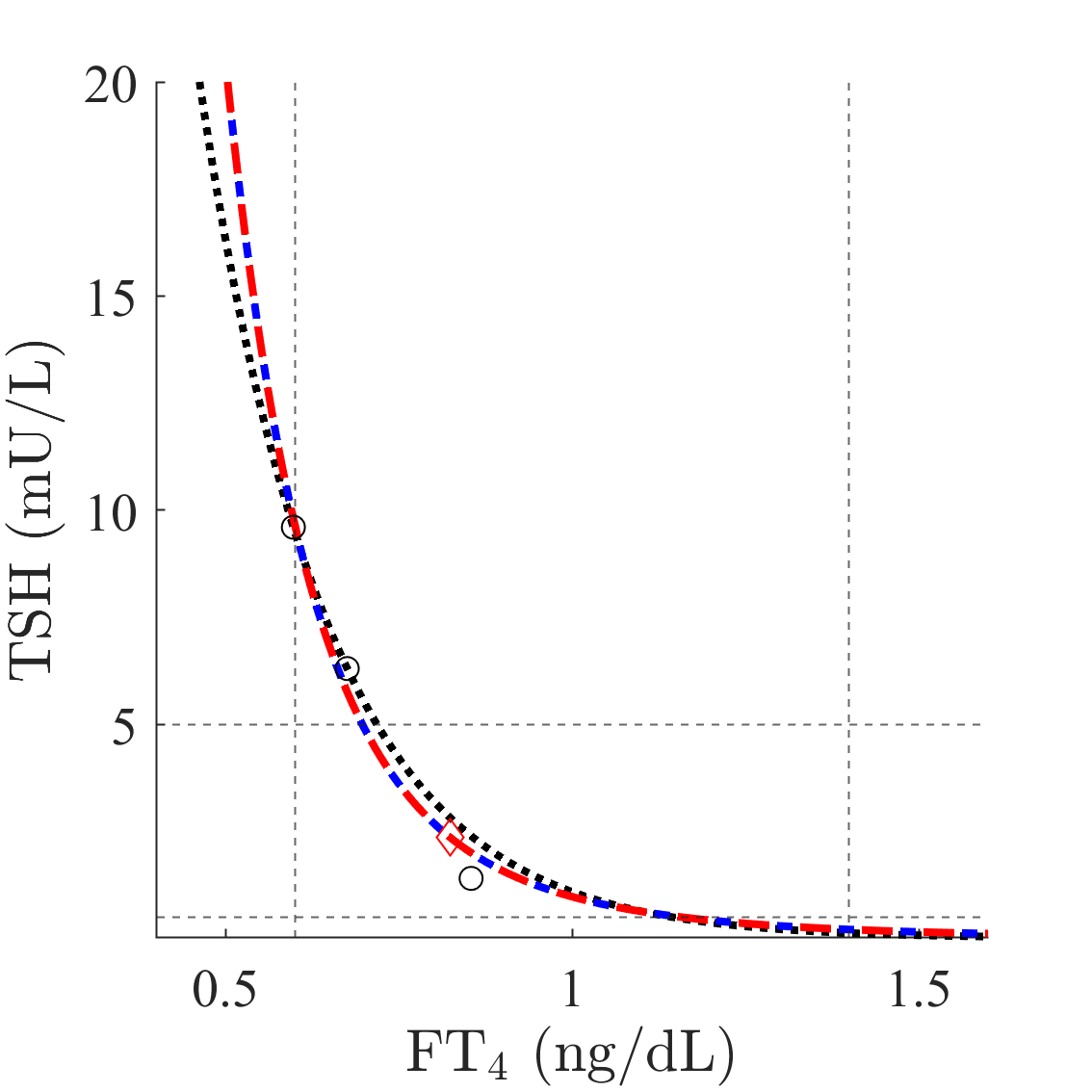}\label{fig:goede4}}
\subfloat[Patient 1IU]{\includegraphics[width=0.33\textwidth, keepaspectratio]{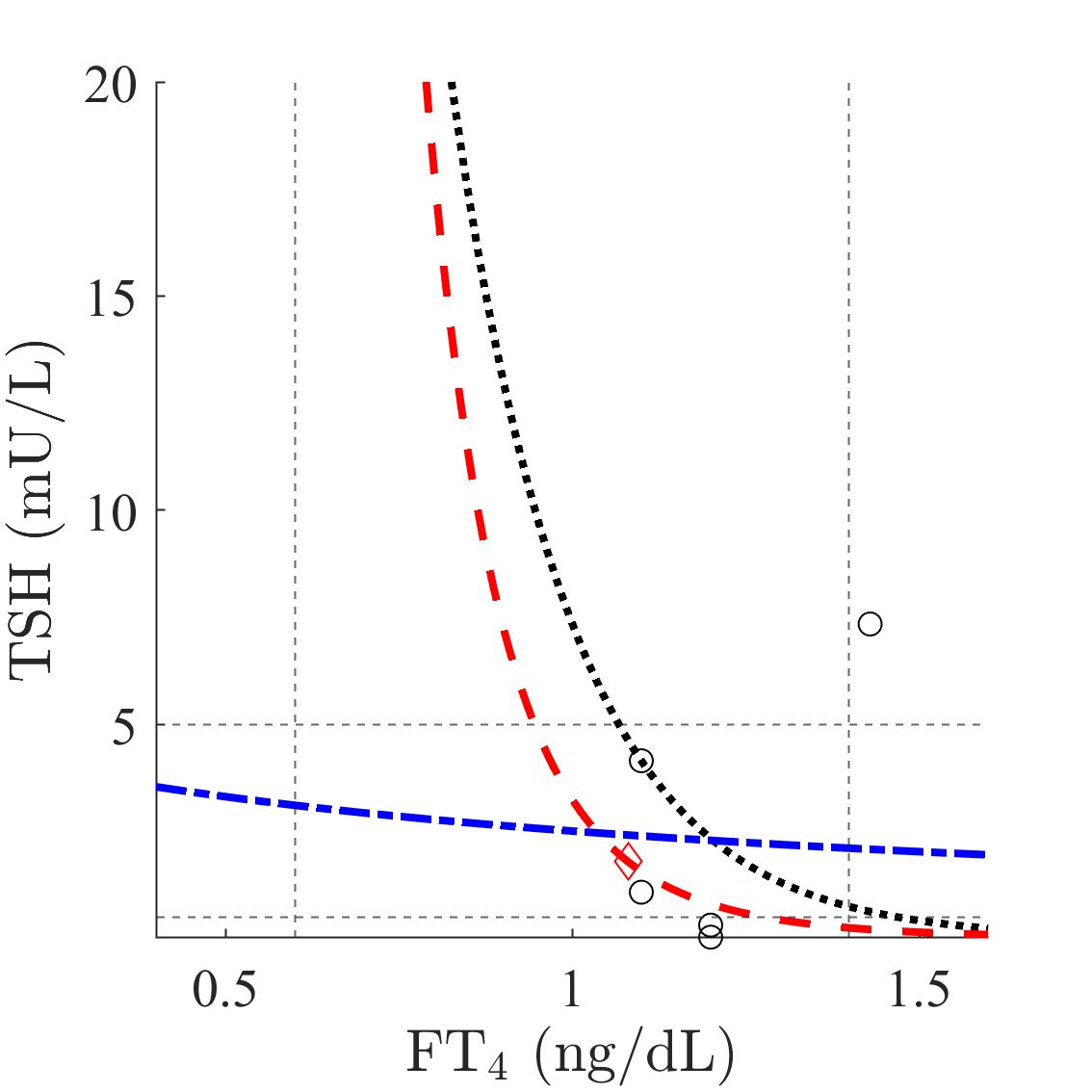}\label{fig:3curvesPatient1}}
\subfloat[Patient 2IU]{\includegraphics[width=0.33\textwidth, keepaspectratio]{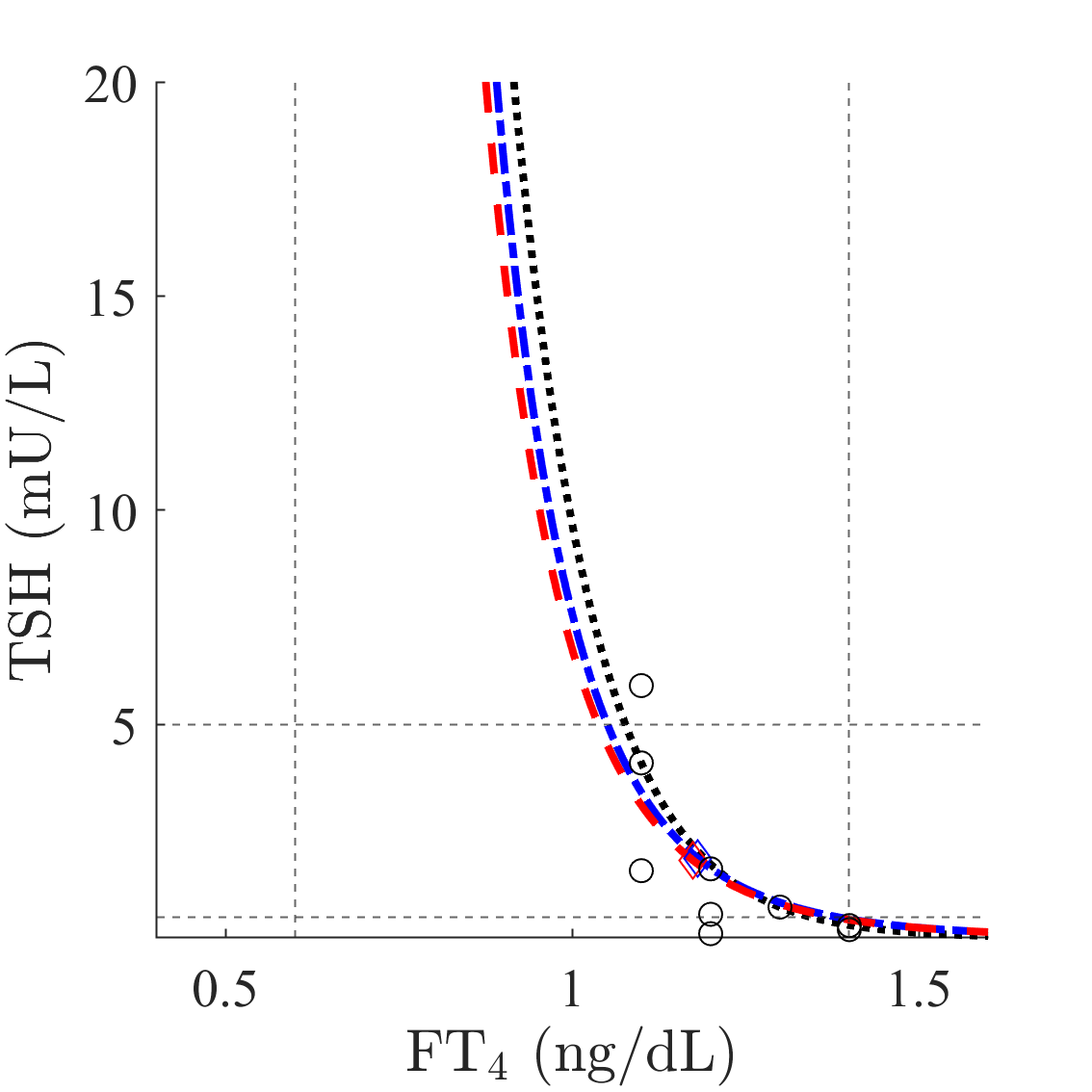}\label{fig:3curvesPatient2}}

\subfloat[Patient 3IU]{\includegraphics[width=0.33\textwidth, keepaspectratio]{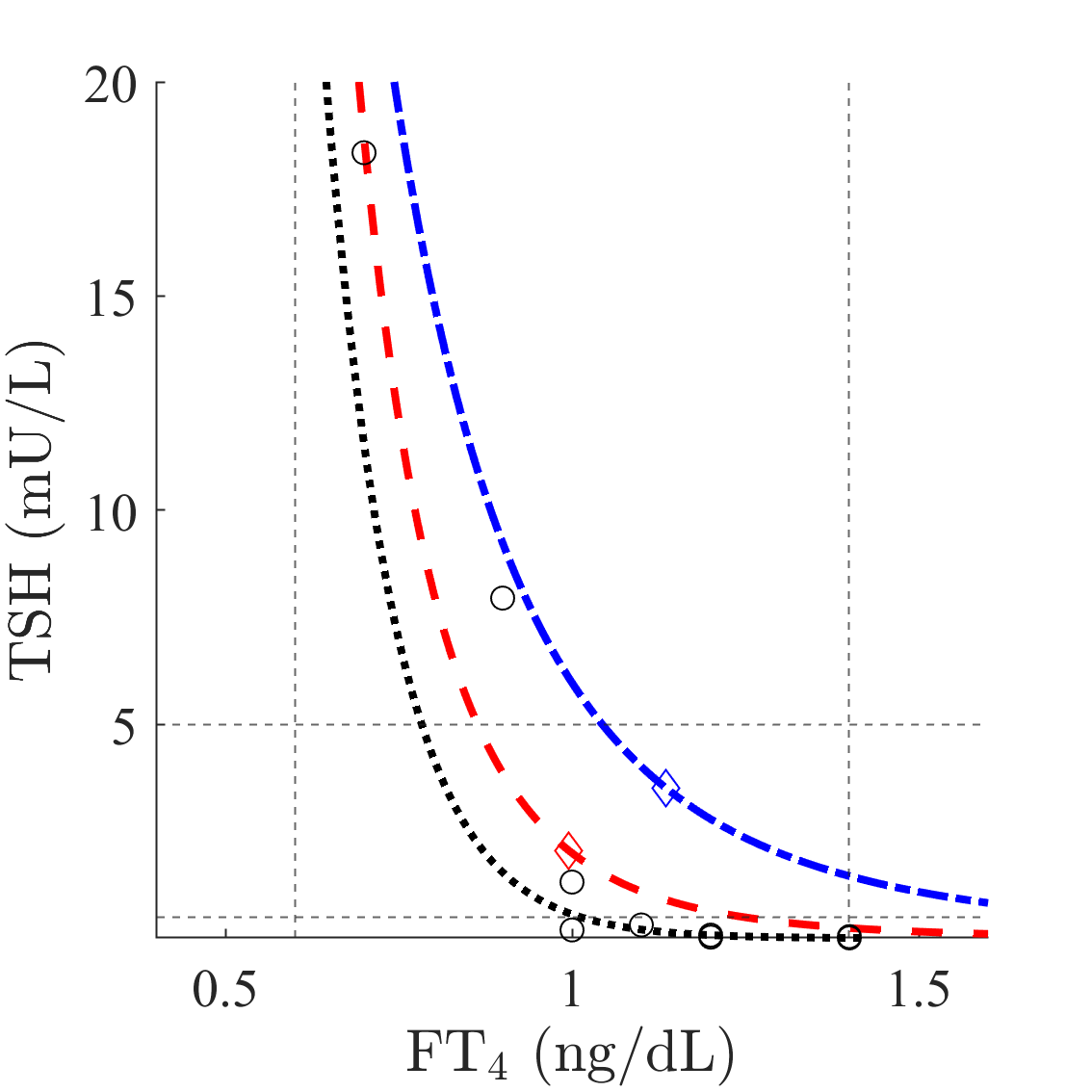}\label{fig:3curvesPatient3}}
\subfloat[Patient 4IU]{\includegraphics[width=0.33\textwidth, keepaspectratio]{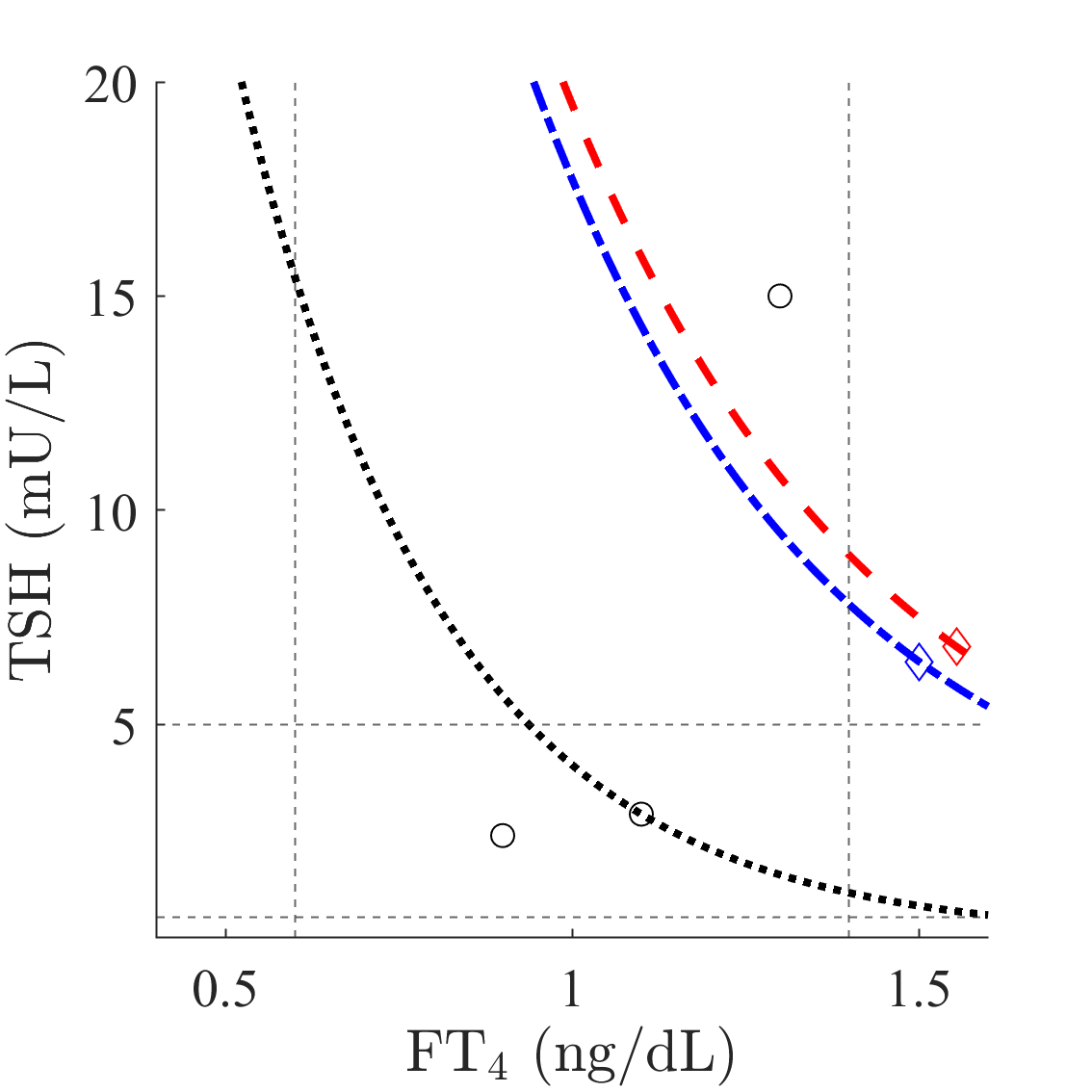}\label{fig:3curvesPatient4}}

\caption []{Comparison of HP curves generated for eight patients using the (i) Goede et. al method, (ii) Improved NLSM and (iii) HLF-NLPS method. Legend for all curves is shown in Subfigure~\ref{fig:goede3}.}
\label{fig:HP_Curves_RealPatients}
\end{figure*}

Figs.~\ref{fig:goede1}--\ref{fig:3curvesPatient4} show the HP curves generated from Patients in Table~\ref{tab:CombinedParams} using the parameters that were obtained from the NLSM and NLPS methods involving $n_1$ estimation. Dashed horizontal and vertical lines indicate the reference ranges for the respective hormone concentrations, which are between 0.5--5 mU/L for TSH, and 0.6--1.4 ng/dL for FT4 \cite{IURefRanges}. The curves generated demonstrate characteristics consistent with the HP curve, as discussed earlier. All three HP curves are presented, as well as the set points found in the estimated HP curves. All eight patients show the characteristic HP curves that we expect to see, with the negative corelation between TSH and FT4. The dashed curve shows the method of Goede et al.~\cite{goede2014novel}, with points selected to generate the best fitting curve by inspection, while the blue and red curves represent the NLSM and NLPS-HLF method respectively. Table~\ref{tab:CombinedParams} show the estimated parameters for \eqref{eq:steadyTSH} used to build the HP curves with the NLSM and NLPS.




\begin{table*}[h]
    \centering
    \caption{Comparison of parameters estimated using NLSM and NLPS methods.}
    \begin{tabular}{ccccccccc}
    \toprule
         & \multicolumn{2}{c}{\textbf{Patient 1IU}} & \multicolumn{2}{c}{\textbf{Patient 2IU}} & \multicolumn{2}{c}{\textbf{Patient 3IU}} & \multicolumn{2}{c}{\textbf{Patient 4IU}} \\
         & NLSM & NLPS & NLSM & NLPS & NLSM & NLPS & NLSM & NLPS \\
    \midrule
    $k_a$ & 11.70 & 37.85 & 22.07 & 25.13 & 999.99 & 13.92 & 999.99 & 999.99 \\
    $k_2$ & 999.99 & 0.00 & 0.00 & 0.00 & 16.15 & 0.01 & 78.24 & 77.74 \\   
    $n_1$ & 0.96 & 7.68 & 7.96 & 7.96 & 4.24 & 6.28 & 3.31 & 3.36 \\
    \bottomrule
    \end{tabular}
    \label{tab:CombinedParams}
\end{table*}

Patients 1G--4G in Figs.~\ref{fig:goede1}--\ref{fig:goede4} had very similar HP curves across all methods of HP curve generation. An exception to this is Patient 1G in Fig.~\ref{fig:goede1}, where the behavior of the NLSM and NLPS-HLF method deviate slightly from the method of~\cite{goede2014novel}. This is likely because of the inclusion of outliers in the estimation methods being considered. 

As discussed above, Patient 1IU in Fig.~\ref{fig:3curvesPatient1} had data that appears flawed, and as such made it difficult to generate a purposeful HP curve. The NLSM method failed to produce a meaningful HP curve, with a set point far below what would be realistic. The method of Goede et al. also had difficulties, as selection of some data points could result in curves that gave a positive corelation between hormone levels, which is biologically implausible. The NLPS-HLF method, in contrast, produced a meaningful HP curve, with a set point almost exactly in the middle of the reference range. Here, we see the significant advantage of the NLPS method in cases when the other techniques fail to produce useful HP curves. Patient 2IU and 3IU in Figures~\ref{fig:3curvesPatient2} and~\ref{fig:3curvesPatient3} respectively show promising results. In both cases, the HP curves and their set points fall well within the reference ranges and closely approximate the patient data.

Patient 2IU is the same data set that produced Fig.~\ref{fig:GoedeBreaker}, which demonstrates the issues with stacked data points. In this instance, both the NLSM and NLPS-HLF methods were able to produce curves relatively similar to one another, and similar to one of the possible curves generated by the method of~\cite{goede2014novel}. The consensus of three different methods towards a single curve indicates a greater likelihood that this generated HP curve is correct.

Patient 3IU demonstrates the strength of the NLPS-HLF method over that of NLSM. In Fig.~\ref{fig:3curvesPatient3}, the NLPS-HLF method generates a curve that better fits the data within the reference range, while the NLSM method does not. This is because the NLSM method prioritizes the greater fit of all the data points to generate its curve, while the NLPS-HLF method will discount the importance of outliers in generating its curve. 

Patient 4IU, shows less certainty in the result. Although both generated HP curves in Fig.~\ref{fig:3curvesPatient4} demonstrate the characteristics of the HP curve, their theorized set points are slightly outside the range boundaries of the reference range. In this case, more patient data would likely be needed within or near the reference range to better fit the curve.

In all cases, the NLPS-HLF method closely reflects the method of Goede et al. both in qualitative and quantitiatve performance. The NLPS-HLF method consistently produces set points that are within the reference ranges of hormones, and generally fits the patient data well. While the NLSM method outperforms the NLPS-HLF method in one case, neither of the set points fall within the reference range, and both are relatively close to one another. 
\section{Discussion}
The set-point of the HP curve for a specific TSH and FT4 value can be a key parameter for an automatic controller to generate the recommended patient-specific medication doses. As opposed to treating individuals with a medication regimen of pills at regular intervals, a controller would automatically detect where an individual’s hormone levels lie on the HP curve and immediately recalculate the required treatment. Previous work by \cite{srinivasan2024model} and \cite{WM2022} present a Model Predictive Control (MPC) based compensator to prescribe daily medication doses for patients with hypothyroidism, with the goal of bringing their hormone levels into healthy ranges. This control strategy takes into account bounds on the dosage a patient requires and is based on a general desired setpoint for the hormones to attain, which is not patient-specific. The inclusion of the HP curve and setpoint presented in this paper as input to the MPC would significantly improve the controller's ability to cater to individual patient needs.

\section{Conclusion} \label{sec:conc}
\subsection{Summary}
Hypothyroidism is a condition that affects up to 10\% of the global population~\citep{chiovato2019}, and treatment is often a difficult task due to the unique nature of each individual's HPT axis. Common practice is to use only TSH to adjust the dose of thyroid replacement in primary hypothyroidism. In this paper, we argue that using both TSH and FT4 values in the replacement may be beneficial in the treatment of hypothyroidism. Our tool in determining optimal values of TSH and FT4 are the HP curves. An HP curve models the hormone concentration relations for every individual, and can be used to determine a person's optimal balance of hormones, which is referred to as the set-point.
An accurate HP curve can assist physicians in determining what the euthyroid condition is for an individual by finding this set-point, and ultimately drive more effective treatment for the condition.

In this paper, we presented a method of dynamically generating HP curves by removing assumptions of constants in the HPT model put forward by~\cite{pandiyan2011mathematical} and adapted by~\cite{bonfantiendocrine}. 
With this, we produced HP curves that were able to account for all patient data that an individual had, thereby generating an HP curve that could be more accurate than the existing method of~\cite{goede2014novel}. 
We then validated these methods using both existing and new patient data, and demonstrated that future use of HP curves and their set points can be used to better drive treatment of patients with hypothyroidism.

\subsection{Open Problems}

This paper does not address the role of T3 hormones in its model, which is another important hormone in the HPT axis. T3 and T4 typically appear at proportional concentrations in the body, so focusing solely on T4 is a modeling simplification. This simplification results in the abstraction of the more active hormone. A model that incorporates this relationship would be more accurate. Future analysis should incorporate a component modeling T3 into the HPT axis model, which can then be used to generate HP curves. Additionally, a comparative study of existing methods for deriving set points is desirable as finding accurate set points is critical to determining optimal treatment strategies.


\bibliography{sn-bibliography}

\end{document}